\documentclass[12pt,reqno]{amsart}
\usepackage{graphicx}
\usepackage{epstopdf}
\RequirePackage{color}
\RequirePackage[colorlinks, urlcolor=my-blue,linkcolor=my-red,citecolor=my-green]{hyperref}
\definecolor{my-blue}{rgb}{0.0,0.0,0.6}
\definecolor{my-red}{rgb}{0.5,0.0,0.0}
\definecolor{my-green}{rgb}{0.0,0.5,0.0}
\definecolor{nicos-red}{rgb}{0.75,0.0,0.0}
\definecolor{light-gray}{gray}{0.6}
\definecolor{really-light-gray}{gray}{0.8}
\definecolor{sussexg}{rgb}{0.0,0.5,0.5}
\definecolor{sussexp}{rgb}{0.5,0.0,0.5}

\usepackage{geometry}                
\usepackage[parfill]{parskip}    
\usepackage{amsmath, amsthm, amssymb}
\usepackage{multirow,url}
\usepackage{verbatim}
\usepackage{amsfonts}
\usepackage{hyperref}
\usepackage{latexsym}
\usepackage{array}
\usepackage{subfig}

\newtheorem{theorem}{Theorem}[section]

\newtheorem{remark}[theorem]{Remark}

\DeclareMathOperator*{\argmax}{arg\,max}

\makeatletter
\newcommand{\addresseshere}{%
  \enddoc@text\let\enddoc@text\relax
}
\makeatother

\begin{document}

\title{Detection of outlying proportions}

\author{Flavio Mignone}

\address{Flavio Mignone \\ Department of Science and Technological Innovation \\ Universit\`a del Piemonte Orientale \\ Viale Teresa Michel, 11, 15121 Alessandria, Italy}
\email{flavio.mignone@uniupo.it}

\author{Fabio Rapallo}

\address{Fabio Rapallo \\ Department of Science and Technological Innovation \\ Universit\`a del Piemonte Orientale \\ Viale Teresa Michel, 11, 15121 Alessandria, Italy}
\email{fabio.rapallo@uniupo.it}





\begin{abstract}
In this paper we introduce a new method for detecting outliers in a set of proportions. It is based on the construction of a suitable two-way contingency table and on the application of an algorithm for the detection of outlying cells in such table. We exploit the special structure of the relevant contingency table to increase the efficiency of the method. The main properties of our algorithm, together with a guide for the choice of the parameters, are investigated through simulations, and in simple cases some theoretical justifications are provided. Several examples on synthetic data and an example based on pseudo-real data from biological experiments demonstrate the good performances of our algorithm.
\end{abstract}



\maketitle

\section{Introduction}
\label{s:intro}

The research presented here comes from a multidisciplinary collaboration concerning a statistical problem in Molecular Biology. Detection of somatic variants, i.e., variants detected only in some cells of a sample, is very important for molecular diagnosis of cancers and to guide therapy decisions. A new experimental approach is based on Next Generation Sequencing (NGS) which has emerged as a powerful tool thanks to several advantages like speed, costs and sensitivity. One main disadvantage of NGS in somatic variant detection is the number of errors introduced by the procedure. This means that many sequences, called reads, show mismatched with respect to the reference sequence: some of this mismatches represent real variants, while others are referred as sequencing errors.

Each variant (true and errors) is weighted with two parameters: the number of times the variant has been detected (number of calls) and sequencing depth, i.e., the number of times the base has been read even without detecting any variant. Basically each variant can be labeled with a proportion between the number of calls and the sequencing depth. Software designed to discriminate between real variants and sequencing errors are referred as variant callers.

In a typical experiment of this kind one collects hundreds of proportions and (a subset of) the available data, virtually, looks like the output in Table \ref{table_ex}. This whole dataset will be analyzed in Sect.~\ref{sims-ex}. Currently available variant callers usually includes more analysis steps and use as input file sequence raw data or alignment files, often in specific format. Note that the data presented here are pseudo-real data. Pseudo-real data have been generated by simulating sequencing data of samples where specific mutation have been introduced in-silico. This kind of data are useful to test tools for NGS data because allow to benchmark them in a controlled environment. Simulated NGS data have thus been processed with a custom analysis pipeline to obtain data shown in Table \ref{table_ex}.

\begin{table}\caption{A portion of a typical table coming from a NGS experiment. First and second columns contain chromosome and position of variant, third and forth columns contain reference and variant nucleotide, fifth and sixth columns contain the number of calls and depth, respectively.}\label{table_ex}
\begin{tabular}{rrrrrr}
\texttt{Chr} & \texttt{Pos} & \texttt{AlleleRef} & \texttt{AlleleVar} & \texttt{NCalls} & \texttt{Depth} \\ \hline
chr17 & 41226495 & T & C & 2 & 5000 \\
chr17 & 41245581 & T & C & 7 & 5000 \\
chr17 & 41203211 & T & C & 3 & 5000 \\
chr17 & 41219580 & T & C & 2 & 2563 \\
chr17 & 41203193 & T & C & 3 & 5000 \\
chr17 & 41245495 & T & C & 1 & 5000 \\
chr17 & 41245628 & T & C & 1 & 5001 \\
chr17 & 41246766 & T & C & 2 & 5000 \\
chr17 & 41234536 & T & C & 1 & 5000 \\
chr17 & 41243472 & T & C & 1 & 2486 \\ \hline
\end{tabular}
\end{table}

The problem is to discriminate between proportions coming from pure noise and proportions coming from a systematic effect. Existing methods for facing this problem can be broadly classified into two groups, described below. A first group of methods, mainly coming from the Bioinformatics literature usually consists of ad-hoc algorithms which use exogenous information (e.g., the calibration of the parameters on a gold standard), and are based on one or more preliminary steps of data selection, see \cite{kroigardetal:16} for a recent review.

From a statistical point of view, there are several techniques for analyzing this kind of data. They are mainly based on pairwise comparison of proportions, see for instance \cite{fleissetal:03}, \cite{newcombe:98}, \cite{mccann|tebbs:07}, \cite{agrestietal:08}, and \cite{nashimotoetal:13}. These methods have two main disadvantages: first, they can be applied to relatively small sets of proportions, while in our framework we need to analyze hundreds or even thousands of proportions simultaneously, and it is not easy to extend the algorithms in the cited literature to large groups of proportions; second, the proportions in a single experiment are usually based on different sequencing depths, and we ask for a method which takes into account this issue. From the data in Table \ref{table_ex}, it is also clear that small proportions are often involved in this kind of problems, and therefore the use of exact non-asymptotic techniques should be preferred.

In this paper, we set our problem in the framework of contingency table analysis and introduce a new method for the detection of outlying proportions based on the search of outlying counts in a contingency table. Therefore, a two-way contingency table with dimensions $2 \times K$ is built using the data, where $K$ is the number of proportions to be analyzed. In this table, we consider the independence model as the base model corresponding to the equality of all the proportions, and we apply a technique for the detection of outlying cells inspired by an algorithm recently introduced in \cite{kuhntetal:14}. That algorithm will be described briefly in the next section. In general, an outlier detection algorithm in the framework of contingency tables considers a base model and then finds the counts which ``appear to be inconsistent with the remainder of that set of data'' \cite{barnett|lewis:94}. The proposed method belongs to the class of two-stage algorithms, where the parameter estimation of the base model and the search of outlying cells is made in two different steps. It has been proved in several papers, see e.g. \cite{shane|simonoff:01} and \cite{kuhnt:04}, that two-stage algorithms outperform one-stage algorithms in terms of robustness and of correct classification rate. Notice that an algorithm of this kind does not need any preliminary data analysis. Moreover, we exploit the special structure of our contingency table to further improve the algorithm.

The paper is organized as follows. In Sect.~\ref{alg-descr} we set up our problem in the framework of contingency table analysis, we recall the basic facts concerning outlier detection in contingency tables, and we introduce the algorithm for finding outlying proportions. In Sect.~\ref{choice} we discuss the issues concerning the choice of the parameters, while in Sect.~\ref{sims-ex} we show the behavior of our algorithm on several simulated scenarios which cover a variety of experimental settings. In addition, the pseudo-real data shown in Table \ref{table_ex} are analyzed. Finally, we conclude with some final remarks and pointers to further research in Sect.~\ref{disc}.

\section{Algorithm for outlier detection} \label{alg-descr}

Let us consider a set of observed proportions
\begin{equation} \label{obsprops}
\hat p_1 = \frac {n_i} {d_i} , \ldots , \hat p_K= \frac {n_k} {d_k}
\end{equation}
coming from $K$ independent experiments. Let $p_1, \ldots , p_K$ be the underlying success probabilities, so that $\hat p_1, \ldots , \hat p_K$ are unbiased estimates of $p_1, \ldots , p_K$, respectively. As discussed in the Introduction, to check the presence of outlying proportions, we define the $2 \times K$ contingency table reported in Table \ref{obstable}, where in the first row there are the counts of the successes and in the second row there are the counts of the failures. Therefore, each column contains the counts from an experiment.
\begin{table}\caption{Observed table.}  \label{obstable}
\begin{tabular}{c|ccccc|c}
 & $1$ & $\cdots$ & $k$ & $\cdots$ & $K$ & \\ \hline
$n$ & $n_1$ & $\cdots$ & $n_k$ & $\cdots$ & $n_K$ & $\sum_k n_k$ \\
$d-n$ & $d_1-n_1$ & $\cdots$ & $d_k-n_k$ & $\cdots$ & $d_K-n_K$ & $\sum_k (d_k-n_k)$ \\ \hline
 $d$ & $d_1$ & $\cdots$ & $d_k$ & $\cdots$ & $d_K$ & $\sum_k d_k$
\end{tabular}
\end{table}
In order to find outlying proportions in (\ref{obsprops}), we move to the search of outlying cells in the contingency table in Table \ref{obstable}. To analyze such table, we first need to choose a sampling scheme. We use here a conditional-Poisson sampling scheme with fixed column sums and thus each $n_k$ is an observation from a random variable $N_k$ with binomial distribution with parameters $d_k$ and $p_k$. In fact, it is well known that when $X, Y$ are independent Poisson random variables with parameters $\lambda_X$ and $\lambda_Y$ respectively, then the conditional distribution of $X$ given $X+Y=x+y$ is binomial with parameters $(x+y)$ and $\lambda_X / (\lambda_X + \lambda_Y)$, see e.g. \cite{shao:03}. For an overview on sampling schemes for contingency tables, see e.g. \cite{bishopetal:07}. For a detailed derivation of the conditional-Poisson sampling scheme in contrast with other schemes, the reader can refer to \cite{haberman:74}.

In this theoretical framework, once a true proportion $p_k$ is estimated with with an estimate $\tilde p_k$, the search for outliers is based on the comparison between the observed count $n_k$ and the outlier region of the relevant binomial distribution. Remember that given a discrete probability distribution with density $f$ and a desired outlying level $\alpha \in (0,1)$, the $\alpha$-outlier region for $f$ is the set ${\mathcal O}_{f,\alpha} = \{n \in {\mathbb N} \ : \ f(n) < k_\alpha\}$, where $k_\alpha$ is the maximum real number such that ${\mathbb P}_f ({\mathcal O}_{f,\alpha}) \leq {\alpha}$. In our study we have also considered a one-tailed version of this definition, because in the problem outlined in the Introduction only unusually large counts are of interest. In such a case, the one-tailed $\alpha$-outlier region for $f$ is the set ${\mathcal O}^+_{f,\alpha} = [\overline n_{\alpha}, +\infty)$, where $\overline n_\alpha$ is the minimum count such that ${\mathbb P}_f ({\mathcal O}^+_{f,\alpha}) \leq {\alpha}$. For a fast computation of $\alpha$-outlier regions in {\tt R} under the most common probability distribution, the package {\tt alphaOutlier} in {\tt R} can be used, see \cite{rehage:15}. In the literature, the outliers are named as types if the observed value exceeds the corresponding expected value, otherwise are named as antitypes. This terminology comes from Configural Frequency Analysis and is now of common use in contingency table analysis, see \cite{kieser|victor:99}, \cite{stemmler:14}, and \cite{rapallo:12}. This is especially useful in this paper, because we will often consider a one-tailed version of the detection algorithm.

Once we have chosen a sampling scheme, the search of outliers in a contingency table needs the choice of a base model, i.e., a model under which we compute the estimate $\tilde p$ to be used in the definition of the outlier region. The equality of all proportions $p_1, \ldots, p_K$ has its counterpart in the assumption of the independence model for the contingency table in Table \ref{obstable} as base model. In fact, consider the table of the expected values in Table \ref{exptable}.
\begin{table}\caption{Table of the expected values under the conditional-Poisson sampling scheme.}  \label{exptable}
\begin{tabular}{c|ccccc|c}
 & $1$ & $\cdots$ & $k$ & $\cdots$ & $K$ & \\ \hline
$n$ & $d_1p_1$ & $\cdots$ & $d_kp_k$ & $\cdots$ & $d_Kp_k$ & $\sum_k d_kp_k$ \\
$d-n$ & $d_1(1-p_1)$ & $\cdots$ & $d_k(1-p_k)$ & $\cdots$ & $d_K(1-p_K)$ & $\sum_k d_k(1-p_k)$ \\ \hline
 $d$ & $d_1$ & $\cdots$ & $d_k$ & $\cdots$ & $d_K$ & $\sum_k d_k$
\end{tabular}
\end{table}
The independence model states that all $2 \times 2$ minors of the table of the expected values vanish, and therefore $p_h = p_k$ for all $h \ne k$, $h,k=1, \ldots, K$.

Notice that in a contingency table each count is a sample of size one from a Poisson (or multinomial) distribution, irrespective of the marginal totals, i.e., irrespective of the actual sample sizes of the experiments. This fact implies that the outlier detection problem must be approached with caution. Moreover, it has a notable effect on estimation and testing. In the framework of contingency tables, it is known that simple algorithms for outlier detection ,e.g. those based on the residuals, suffer from masking and swamping. Due to the fact that outlying cells contribute to the actual value of the probability estimates, biased estimates under the base model may appear, and the presence of one or more outliers may be not recognized by the procedure (masking), and in extreme cases it is also possible that the role of outlying and inlying counts is switched (swamping), yielding totally inaccurate conclusions. For details on this issue, see \cite{davies|gather:93}, \cite{kuhnt|pawlitschko:05}, and \cite{kuhnt:10}.

To detect outliers in a table like Table \ref{obstable}, we adopt a strategy with the same philosophy of the algorithm based on minimal patterns introduced in \cite{kuhntetal:14}. Let us recall briefly that algorithm. In the general framework of contingency tables, minimal patterns are subsets of cells that lead to a non-singular design matrix under the base model, and therefore allow the estimation of all cell probabilities. To prevent masking and swamping, once a minimal pattern has been chosen, only the cells outside the minimal pattern are tested for the presence of outliers. This procedure is repeated for all possible minimal patterns and a cell is finally declared as outliers when it is recognized in the $\alpha$-outlier region more than half the times.

Given a contingency table with $N$ cell counts $y_j$, $j=1, \ldots N$, the entries are assumed to be realizations of random variables $Y_j$, $j=1,...,N$, from a log-linear Poisson model. The structural component of a log-linear model can be written in the form:
\[
 E(Y_j) =  \exp(x_j' \beta)  , \ j = 1,...,N \, ,
\]
where $x_j$ is the $j^{th}$ column of the design matrix $X \in {\mathbb R}^{p \times N}$ of the model and $\beta\in \mathbb R^p$ is the vector containing the unknown parameters. Here we assume that the model is parameterized in such a way that $X$ is full-rank. Notice that in this notation each row of $X$ represents an unknown parameter of the model, while each column represent a cell of of the contingency table. The maximum likelihood estimator of $\beta$ is given by:
\begin{equation} \label{MLE-est}
\widehat{\beta}^{ML}
={\argmax}_{\beta\in\mathbb R^p}\left(\sum_{j=1}^N\left(
Y_j\, x_j'\beta-\exp({x_j'\beta})\right)\right) \, .
\end{equation}
In this context, a minimal pattern is a subset with at least $N/2$ cells, such that the matrix $X$ restricted to the chosen cells is full rank, and with minimal cardinality to fulfill both conditions. Let ${\mathcal M}$ be the set of all minimal patterns for the model matrix $X$.

Given the set $\mathcal W$ containing all the $W$ minimal patterns, the outlier detection algorithm based on minimal patterns (OMPC) under the Poisson sampling scheme can be summarized as follows:
\begin{enumerate}
\item for each $w \in {\mathcal W}$:
\begin{itemize}
\item[(1a)] Compute the parameter estimates
\[
\widehat{\beta}_w^{ML} = {\argmax}_{\beta\in\mathbb R^p}\left(\sum_{j \in w} \left(
Y_j\, x_j'\beta-\exp({x_j'\beta})\right)\right)
\]
as in Eq. (\ref{MLE-est}), but based on the minimal pattern $w$;

\item[(1b)] For $j=1, \ldots N$, $j \notin w$, compute the outlier region $out(\alpha, Pois(\hat{m}^w_j))$ for $m^w_j$ based on          $\exp(x_j'\widehat{\beta}_w^{ML})$;
\end{itemize}
\item Define $r_j$ as the absolute frequency of cell $j$ not contained in a minimal pattern;

\item Compare the observed count in the cell $y_j$ with the outlier regions from item (1b). If {$\# (y_j \in out(\alpha,Pois(\hat{m}^w_j)), j \notin w) > r_j/2$ } then $y_j$ is declared to be an outlier.
\end{enumerate}

In cases where $W$ becomes large and the enumeration of all minimal patterns is not feasible, it is possible to introduce a standard Monte
Carlo approximation in the algorithm. For details on the choice of the parameters and for several examples on both simulated and experimental datasets, see \cite{kuhntetal:14}.

For analyzing a $2 \times K$ table under the independence model the algorithm above would need minimal patterns with $K+1$ cells, and this yields estimates of the proportions highly influenced by the presence of outliers. Indeed, using the Poisson model for the contingency table in Table \ref{obstable}, we must consider $K+1$ parameters: one parameter for the overall mean, one parameter for the row effect, and $K-1$ parameters for the column effect. A valid minimal pattern is formed by the counts in the second row of the observed table plus one count in the first row, and in such a case an outlier in the chosen numerator would produce a strong bias in the estimation of the parameter governing the row effect.

Moreover, the algorithm in \cite{kuhntetal:14} is designed for small- to moderate-sized tables, and no study is available when it is applied to large tables, while in our settings it is customary to have values such as $K=100$ or more, leading to tables with $200$ cells or more. Therefore, we propose a different approach to be used in our framework. First, we take advantage of the different sampling scheme chosen above, namely the conditional-Poisson scheme, and thus we do not consider minimal patterns of the whole table, but we simply choose a sub-table with $H$ columns in order to estimate the common binomial parameter $p = p_1 = \ldots p_K$ under the independence model. Secondly, we use the binomial distribution instead of the Poisson distribution. Third, we choose different parameters in the algorithm in order to encompass the application also to large sets of observed proportions. We will consider this last issue in the next sections.

Let $T$ be the contingency under analysis, in the format illustrated in Table \ref{obstable}, and let ${\mathcal S}_H$ be the set of all minimal patterns with $H$ columns. Each ${\mathcal H} \in {\mathcal S}_H$ consists of $H$ columns of the table $T$ chosen with a uniform sampling without replacement. The proposed algorithm proceeds as follows:
\begin{enumerate}
\item For each $b=1, \ldots, B$:
\begin{itemize}
\item[1a] Choose a minimal patterns ${\mathcal H} \in {\mathcal S}_H$;

\item[1b] Estimate the common proportion $p$ with
\begin{equation} \label{est-min-pat}
\tilde p = \frac {\sum_{h \in {\mathcal H}} n_h } {\sum_{h \in {\mathcal H}} d_h} \, .
\end{equation}

\item[1c] For each numerator $N_k$ with $k \in \{1, \ldots, K\} \setminus {\mathcal H}$, compare $n_k$ with the $\alpha$-outlier region of a binomial distribution with parameters $d_k$ and $\tilde p$.
\end{itemize}

\item For each $k \in \{1, \ldots , K\}$, let $C_k$ be the number of checks made at item $1c$, and let $S_k$ be the number of positive checks. If the ratio $S_k/C_k$ exceeds a fixed number $r$, then the $k$-th proportion is declared to be an outlier.
\end{enumerate}

In the case of proportions, the algorithm is presented directly in its Monte-Carlo version, because the number of possible minimal patterns is large already in small cases. For instance, when we analyze a set of $20$ proportions, the number of minimal patterns is $184,756$, and a Monte-Carlo approximation with $B=1,000$ yields very stable results.

\begin{remark}
The algorithm described here can be applied in both one-sided and two-sided problems, while pairwise comparisons of proportions do not. This issue is particularly relevant in our application, since usually only one-sided problems arise.
\end{remark}

The algorithm above depends on three parameters: the level $\alpha$, the size of the minimal patterns $H$, and the critical ratio $r$ in the last item. While the level $\alpha$ may be fixed a priori, the parameters $H$ and $r$ must be chosen in order to control the sensitivity and/or the specificity of the algorithm. In the next sections, we discuss this issue with some computations in simple settings, and through a simulation study in more complex scenarios. In all cases, we will consider only the one-tailed problem, because it is of prominent interest in our motivating application.

To compute the Mean Squared Error ${\mathrm{MSE}}(\tilde P)$, we need to specify the nature of the outlying cells. Here we assume the outlying counts $n_i$ as fixed. Thus, when there are $NO$ outliers, we assume
\[
N_i = n_i \qquad \qquad \mbox{ for } \ i = 1, \ldots , NO
\]
and
\[
N_i \sim {\mathrm{Bin}}(d_i,p) \qquad \qquad \mbox{ for } \ i = NO+1, \ldots , K \, .
\]
For a minimal pattern ${\mathcal H}$, the bias is
\begin{equation} \label{bias-tildep}
\Delta({\tilde P})_{H} = \frac {\sum_{i \in {\mathcal H}, i = 1}^{NO} (n_i-d_i p)  } {\sum_{i \in {\mathcal H}} d_i}
\end{equation}
and the variance is
\begin{equation} \label{var-tildep}
\mathrm{var}({\tilde P})_H = \frac {p(1-p) \sum_{i \in {\mathcal H}, i =NO+1}^{K} d_i} {\left(\sum_{i \in \mathcal H} d_i\right)^2}\ , .
\end{equation}
Combining Eq. (\ref{bias-tildep}) and (\ref{var-tildep}) we obtain
\begin{equation} \label{MSE-tildep}
{\mathrm{MSE}}(\tilde P) = \frac 1 {\#{\mathcal S}} \sum_{{\mathcal H} \in {\mathcal S}} \Delta({\tilde P})_H^2 + \mathrm{var}({\tilde P})_H \, ,
\end{equation}
where $\#{\mathcal S}_H$ denotes the cardinality of ${\mathcal S}_H$.

Notice that the ${\mathrm{MSE}}$ in Eq. (\ref{MSE-tildep}) needs a prior knowledge of the number and the position of the outliers in the table, and therefore it cannot be used to actually evaluate the accuracy of the proposed algorithm. To assess the performance of the algorithm in simulated scenarios, where the outliers are known, we have used sensitivity (sens) and specificity (spec), i.e., the proportion of correctly classified outliers and the proportion of correctly classified inliers, respectively.

\section{Choice of the parameters} \label{choice}

In this section, we illustrate how to approach the problem of the choice of the parameters in the proposed algorithm. In simple cases (e.g., when there is only one outlying proportion), we are able to evaluate the performance of the algorithm with theoretical arguments, while in general we perform a simulation study. Through such study, we show that the algorithm exhibits good performances in terms of sensitivity and specificity in a large spectrum of choices of the level $\alpha$, and therefore it represents a first illustration of the validity of the algorithm. Further simulations and examples will be presented in the next section.

\subsection{Choice of $\alpha$}

The choice of the level $\alpha$ depends strongly on the user's preferences. Usually, values of $\alpha$ ranging from $10^{-4}$ and $10^{-3}$ are a reasonable choice in the case of count data, see for instance \cite{kuhntetal:14}. When the number of proportions under investigation become very large, one may choose to decrease $\alpha$ up to $10^{-5}$ or even less, while for small sets of proportions one may decide to increase $\alpha$ up to $10^{-2}$. To show that our algorithm works well for values of $\alpha$ in a reasonable range, and to show that the parameter $\alpha$ basically controls the sensitivity and the specificity of the algorithm, we have performed a preliminary simulation study whose results are reported in Table \ref{simalpha}. The algorithm is tested in different scenarios:
\begin{itemize}
\item one outlier (type) out of $20$ proportions, with denominators $d_h=100$, and with three different choices of $p$;

\item three outliers (types) out of $100$ proportions, with denominators $d_h=100$, and with three different choices of $p$;

\item three outliers (both types and antitypes) out of $100$ proportions, with denominators $d_h=100$, and with three different choices of $p$;

\item three outliers (both types and antitypes) out of $100$ proportions, with denominators $d_h=1,000$, and with three different choices of $p$;
\end{itemize}
For a given level $\alpha$, the outliers are generated from the right tail with weight $\alpha$ of the relevant binomial distribution in the one-tail case, while in the two-tail case they are generated from the right or from the left tail with probability $1/2$ each, when this makes sense, i.e, when the zero count does not belong to the inlier region; when the zero count belongs to the inlier region, the left tail is ignored. In all cases we also exclude the smallest count in the right tail and the largest count in the left tail of the outlier region. This prevents from a border effect which yields smaller values of sensitivity and leads to less clear results. We will discuss this issue in the next paragraph. At this stage, the other parameters are fixed at $H=0.5$ and $r=0.5$. In each scenario we generate $1,000$ random data sets.

\begin{table}  \caption{Sensitivity and specificity of the proposed algorithm for several values of $\alpha$ in different scenarios. $NO$ is the number of outliers in each configuration.} \label{simalpha}
\rotatebox{90}
{
\begin{tabular}{ccccc|ccccccccccc}
$K$ & $NO$ & $d_h$ & p & & \multicolumn{11}{c}{$\alpha$} \\
    &            &       &   & & $10^{-6}$ & $5\cdot 10^{-6}$ & $10^{-5}$ & $5\cdot 10^{-5}$ & $10^{-4}$ & $5\cdot 10^{-4}$ & $10^{-3}$ & $5\cdot 10^{-3}$ & $10^{-2}$ & $5\cdot 10^{-2}$ & $10^{-1}$ \\ \hline
\multicolumn{16}{c}{One-tailed scenarios} \\
20 & 1 & 100 & 0.01 & sens & 0.929 & 0.988 & 0.941 & 1.000 & 0.975 & 1.000 & 0.994 & 0.984 & 1.000 & 1.000 & 1.000 \\
   &   &     &      & spec & 1.000 & 1.000 & 1.000 & 1.000 & 1.000 & 1.000 & 1.000 & 0.999 & 0.997 & 0.983 & 0.948 \\
20 & 1 & 100 & 0.05 & sens & 0.960 & 0.966 & 0.898 & 0.953 & 0.988 & 0.913 & 0.988 & 0.950 & 0.997 & 1.000 & 1.000 \\
   &   &     &      & spec & 1.000 & 1.000 & 1.000 & 1.000 & 1.000 & 1.000 & 0.999 & 0.998 & 0.995 & 0.974 & 0.934 \\
20 & 1 & 100 & 0.10 & sens & 0.975 & 0.919 & 0.972 & 0.894 & 0.984 & 0.950 & 0.929 & 0.929 & 0.994 & 0.975 & 0.984 \\
   &   &     &      & spec & 1.000 & 1.000 & 1.000 & 1.000 & 1.000 & 1.000 & 1.000 & 0.997 & 0.994 & 0.967 & 0.937 \\
100& 3 & 100 & 0.01 & sens & 0.933 & 1.000 & 0.980 & 1.000 & 0.998 & 1.000 & 1.000 & 1.000 & 1.000 & 1.000 & 1.000 \\
   &   &     &      & spec & 1.000 & 1.000 & 1.000 & 1.000 & 1.000 & 1.000 & 1.000 & 0.999 & 0.997 & 0.982 & 0.945 \\
100& 3 & 100 & 0.05 & sens & 0.990 & 0.998 & 0.922 & 0.989 & 1.000 & 0.975 & 1.000 & 0.998 & 1.000 & 1.000 & 1.000 \\
   &   &     &      & spec & 1.000 & 1.000 & 1.000 & 1.000 & 1.000 & 1.000 & 1.000 & 0.998 & 0.996 & 0.972 & 0.938 \\
100& 3 & 100 & 0.10 & sens & 0.996 & 0.937 & 0.997 & 0.974 & 1.000 & 0.998 & 0.995 & 0.990 & 1.000 & 1.000 & 1.000 \\
   &   &     &      & spec & 1.000 & 1.000 & 1.000 & 1.000 & 1.000 & 1.000 & 1.000 & 0.997 & 0.995 & 0.968 & 0.929 \\     \hline \hline
\multicolumn{16}{c}{Two-tailed scenarios} \\
100& 3 & 100 & 0.05 & sens & 0.984 & 0.994 & 0.917 & 0.991 & 1.000 & 0.991 & 1.000 & 0.994 & 1.000 & 0.998 & 1.000 \\
   &   &     &      & spec & 1.000 & 1.000 & 1.000 & 1.000 & 1.000 & 1.000 & 1.000 & 0.998 & 0.994 & 0.961 & 0.915 \\
100& 3 & 100 & 0.10 & sens & 0.998 & 0.960 & 1.000 & 1.000 & 1.000 & 0.976 & 1.000 & 0.998 & 1.000 & 1.000 & 0.991 \\
   &   &     &      & spec & 1.000 & 1.000 & 1.000 & 1.000 & 1.000 & 1.000 & 0.999 & 0.996 & 0.991 & 0.956 & 0.904 \\
100& 3 & 100 & 0.20 & sens & 1.000 & 0.964 & 0.998 & 1.000 & 0.986 & 0.995 & 1.000 & 1.000 & 1.000 & 1.000 & 1.000 \\
   &   &     &      & spec & 1.000 & 1.000 & 1.000 & 1.000 & 1.000 & 1.000 & 0.999 & 0.995 & 0.991 & 0.951 & 0.917 \\
100& 3 & 1000& 0.05 & sens & 0.943 & 0.975 & 0.968 & 0.984 & 0.979 & 0.976 & 0.994 & 0.970 & 0.995 & 0.980 & 0.994 \\
   &   &     &      & spec & 1.000 & 1.000 & 1.000 & 1.000 & 1.000 & 0.999 & 0.999 & 0.995 & 0.991 & 0.953 & 0.902 \\
100& 3 & 1000& 0.10 & sens & 0.940 & 0.949 & 0.968 & 0.944 & 0.966 & 0.962 & 0.978 & 0.989 & 0.962 & 0.983 & 0.984 \\
   &   &     &      & spec & 1.000 & 1.000 & 1.000 & 1.000 & 1.000 & 0.999 & 0.999 & 0.995 & 0.989 & 0.952 & 0.905 \\
100& 3 & 1000& 0.20 & sens & 0.917 & 0.946 & 0.919 & 0.930 & 0.951 & 0.906 & 0.951 & 0.936 & 0.975 & 0.969 & 0.967 \\
   &   &     &      & spec & 1.000 & 1.000 & 1.000 & 1.000 & 1.000 & 0.999 & 0.999 & 0.995 & 0.990 & 0.950 & 0.899 \\ \hline \hline
\end{tabular}
}
\end{table}

The results in Table \ref{simalpha} show that in all experimental settings the algorithm reaches very high values of sensitivity and specificity. In the one-tailed scenarios, the sensitivity is greater than $95\%$ in almost all cases and tends to increase for higher values of $\alpha$, as expected. Also the the specificity is greater than $95\%$ in the great majority of cases and tends to decrease for higher values of $\alpha$. In the two-tailed scenarios, the results are almost the same, although in one case the specificity is less than $90\%$, but this happens when $\alpha$ is fixed to $10^{-1}$, a rather extreme value of $\alpha$ in this framework. In all scenarios, both sensitivity and specificity are greater than $98\%$ in almost all settings when restricting to $\alpha$ between $10^{-4}$ and $10^{-3}$, the most common values of $\alpha$ when searching possible outliers.
Finally, the algorithm yields results with the same accuracy in both one- and two-tailed scenarios. Further simulations with varying denominators $d_i$'s will be discussed in the next section.

\subsection{Choice of $r$}

As a first choice for the parameter $r$, we opt for $r=0.5$, and this choice can be justified as follows. Consider the special case of a set of proportions with only one outlier, say the first proportion. For sake of simplicity we consider the one-tailed version of the algorithm with $p<0.5$, but the same argument applies to the other cases. Since there is only one outlier, the outlyingness of the first count $n_1$ is based on minimal patterns not containing $n_1$. Therefore, the outlier region is based on an unbiased estimate $\tilde p$ of $p$ in Eq. (\ref{est-min-pat}). Denote with $\tilde P$ the corresponding estimator. Under the large-sample approximation, the distribution of $\tilde P$ is approximately Gaussian with mean $p$, and
\[
{\mathbb P}(\tilde P > p) = 0.5 \, , \qquad \qquad {\mathbb P}(\tilde P < p) = 0.5 \, .
\]
In the worst case, when the outlying count $n_1$ lies on the boundary of the outlier region, each check in step $(1c)$ of the proposed algorithm yields a random variable $S_k$ with Bernoulli distribution with parameter $0.5$, and therefore $0.5$ is the expected value of the ratio $S_k/C_k$. This fact implies that the outlier is classified correctly with probability $0.5$ and the sensitivity is $0.5$. However, as soon as the outlier is not on the boundary of the outlier region, the sensitivity goes to $1$ as the number of Monte Carlo replicates $W$ goes to infinity. Regarding the specificity, each inlier is correctly classified in each step $(1c)$ of the algorithm with probability $(1-\alpha)$ and thus the specificity goes to $1-\alpha$ as $W$ goes to infinity also in the worst case of an outlying count on the boundary of the outlier region.

\begin{remark}
Since the computation of the estimate $\tilde p$ in each step has in its denominator the sum of the $d_i$'s in the minimal pattern, the large sample approximation for $\tilde P$ is reasonable also when a small number of proportions is analyzed.
\end{remark}

When the true number of outliers is greater than 1, sensitivity and specificity can not be computed directly and we revert again to simulations.

\subsection{Choice of $H$}

The philosophy underlying the method of minimal patterns is to estimate the parameters, i.e., the common probability of success $p$ in our setting, through an unbiased estimator. Therefore, the choice of $H$ would contrast these two opposite requirements: on one side, small values of $H$ lead to a large probability that the minimal patterns do not contain outliers and thus the estimator $\tilde P$ is unbiased; on the other side, large values of $H$ yield estimators with smaller variance. Notice that an evaluation of the probability of having outlier-free minimal patterns would require the prior knowledge of the number of outliers, and therefore such evaluation would have no practical meaning. Following \cite{kuhntetal:14}, we set $H=0.5$, and this value has been used in all our examples and simulations.

\section{Simulations and a case study} \label{sims-ex}

Since in our preliminary simulations in Sect.~\ref{choice} we have shown that the algorithm yields very good values of sensitivity and specificity in a wide range of choices of the parameter $\alpha$, we concentrate now on the most usual values of $\alpha$ (i.e., between $10^{-4}$ and $10^{-3}$), and we simulate some situations where the values of $d_i$ are not constant. We consider values of $d_i$ with different orders of magnitude in the same table. More precisely, the denominators are chosen to be $d_h=100$ or $d_=1,000$. The results are reported in Table \ref{simden}. The algorithm is tested in different scenarios:
\begin{itemize}
\item one outlier (type) out of $20$ proportions, with three different choices of $p$;

\item three outliers (types) out of $100$ proportions, with three different choices of $p$;

\item ten outliers (types) out of $500$ proportions, with three different choices of $p$;

\item three outliers (both types and antitypes) out of $100$ proportions, with four different choices of $p$;

\item ten outliers (both types and antitypes) out of $500$ proportions, with four different choices of $p$;
\end{itemize}
The outliers are generated with the same rule as in Sect.~\ref{choice}. As mentioned in the Introduction, the specific routines for NGS analysis use as input file sequence raw data or alignment files, and thus they can not be compared with our algorithm on simulated data.

All the results in Table \ref{simden} show that the proposed algorithm has good performances also in these new scenarios, and its behavior is not affected by the presence of different orders of magnitude in the $d'i$'s.

\begin{table}  \caption{Sensitivity and specificity of the proposed algorithm in several scenarios with non-constant $d_i$'s. $NO$ is the number of outliers in each configuration.} \label{simden}
\begin{tabular}{cccc|ccc}
$K$ & $NO$ & p & & \multicolumn{3}{c}{$\alpha$} \\
    &      &   & & $10^{-4}$ & $5\cdot 10^{-4}$ & $10^{-3}$ \\ \hline
\multicolumn{7}{c}{One-tailed scenarios} \\
20 & 1 & 0.01 & sens & 0.944 & 0.963 & 0.969  \\
   &   &      & spec & 1.000 & 1.000 & 1.000  \\
20 & 1 & 0.05 & sens & 0.919 & 0.910 & 0.950  \\
   &   &      & spec & 1.000 & 1.000 & 1.000  \\
20 & 1 & 0.10 & sens & 0.898 & 0.879 & 0.913  \\
   &   &      & spec & 1.000 & 1.000 & 0.999  \\
100& 3 & 0.01 & sens & 0.986 & 0.988 & 0.979  \\
   &   &      & spec & 1.000 & 1.000 & 1.000  \\
100& 3 & 0.05 & sens & 0.888 & 0.909 & 0.974  \\
   &   &      & spec & 1.000 & 1.000 & 1.000  \\
100& 3 & 0.10 & sens & 0.909 & 0.900 & 0.902  \\
   &   &      & spec & 1.000 & 1.000 & 1.000  \\
500& 10& 0.01 & sens & 0.993 & 1.000 & 1.000  \\
   &   &      & spec & 1.000 & 1.000 & 0.997  \\
500& 10& 0.05 & sens & 0.921 & 0.981 & 0.999  \\
   &   &      & spec & 1.000 & 1.000 & 0.999  \\
500& 10& 0.10 & sens & 0.973 & 0.919 & 0.946  \\
   &   &      & spec & 1.000 & 1.000 & 0.999  \\ \hline \hline
\multicolumn{7}{c}{Two-tailed scenarios} \\
100& 3 & 0.05 & sens & 0.979 & 0.981 & 0.992  \\
   &   &      & spec & 1.000 & 1.000 & 0.999  \\
100& 3 & 0.10 & sens & 0.950 & 0.964 & 0.974  \\
   &   &      & spec & 1.000 & 1.000 & 0.999  \\
100& 3 & 0.20 & sens & 0.965 & 0.932 & 0.966  \\
   &   &      & spec & 1.000 & 1.000 & 0.999  \\
100& 3 & 0.50 & sens & 0.935 & 0.964 & 0.960  \\
   &   &      & spec & 1.000 & 1.000 & 0.999  \\
500& 10& 0.05 & sens & 0.993 & 0.996 & 1.000  \\
   &   &      & spec & 1.000 & 1.000 & 0.999  \\
500& 10& 0.10 & sens & 0.994 & 0.998 & 0.999  \\
   &   &      & spec & 1.000 & 1.000 & 0.999  \\
500& 10& 0.20 & sens & 0.995 & 0.986 & 0.995  \\
   &   &      & spec & 1.000 & 1.000 & 0.999  \\
500& 10& 0.50 & sens & 0.990 & 0.996 & 0.995  \\
   &   &      & spec & 1.000 & 1.000 & 0.999  \\ \hline \hline
\end{tabular}
\end{table}

Finally, we consider the dataset illustrated in the introduction, with an excerpt in Table \ref{table_ex}. The whole dataset consists of $3572$ lines and the data are simulated in order to have $26$ outlying proportions. Using our algorithm with two different values of $\alpha$ and we the other parameters set as discussed in Sect. \ref{choice}, we obtain the following results:
\begin{itemize}
\item with $\alpha = 10^{-3}$, there are only $3$ errors, namely $3$ false positives, and thus sens$=1$ and spec$=0.999$.

\item with $\alpha = 10^{-4}$, all the $3572$ are classified correctly. Both sensitivity and specificity are equal to 1.
\end{itemize}

\section{Discussion} \label{disc}
\label{s:discuss}

The analysis of simulated scenarios and the pseudo-real data example show that the proposed algorithm has high values of sensitivity and specificity in all the considered settings. Among the main features of our method we would like to emphasize that: it can be applied when a large number of proportions needs to be analyzed; it does not require any calibration on gold standards; it takes into account the presence of different depths in the same dataset. Moreover, we have considered here only the independence model as base model, but the algorithm can be easily extended to other models, such as logistic regression models.

The application of our algorithm together with the comparison of its performances with those of other methods used in NGS routines is currently in progress and will be included in a more specialized paper addressed to a partially different audience. However, the first preliminary results in this direction are promising. Another future direction of this research will consist in the study of this type of algorithms within the framework of hypothesis testing as in \cite{rapallo:12}, using algebraic tools for exact testing.



\section*{Acknowledgements}

The authors thank Dr. Francesco Favero (University of Piemonte Orientale) for some helpful suggestions and for having generated simulated NGS data. This research is original and has a financial support of the Universit\`a del Piemonte Orientale. \vspace*{-8pt}


%

\bibliographystyle{alpha}
\bibliography{MR-bib}

\end{document}